\documentclass[aip,reprint]{revtex4-1}
\pdfoutput=1 

\usepackage{natbib}
\usepackage{graphicx}
\usepackage{subfig}
\usepackage{amssymb}
\usepackage{amsfonts}

\begin{document}

\title{Technical Note: Radiotherapy dose calculations using GEANT4 and the Amazon Elastic Compute Cloud}

\author{C M Poole}
\email{christopher.poole@qut.edu.au}
\affiliation{Discipline of Physics, Faculty of Science and Technology\\
    Queensland University of Technology, Brisbane, Australia}
\author{I Cornelius}
\author{J V Trapp}
\author{C M Langton}

\date{\today}

\begin{abstract}
Cloud computing allows for vast computational resources to be leveraged quickly and easily in bursts as and when required.
Using the Amazon Elastic Compute Cloud and the Amazon Simple Storage Solution, we describe a technique that allows for Monte Carlo radiotherapy dose calculations to be performed using GEANT4 and executed in the cloud.
Simulation cost and completion time was evaluated as a function of instance count using compute instances acquired via biding on the Elastic Compute Cloud spot market.
Bidding for instances on the instance spot market was found to be 35-60\% of the cost of on-demand instances of the same type.
Using the technique, we demonstrate the potential usefulness of cloud computing as a solution for rapid Monte Carlo simulation for radiotherapy dose calculation.
\end{abstract}
\maketitle

\section{Introduction}
GEANT4 is a C++ toolkit for the simulation of particle transport though geometry, and is used widely in the field of high energy physics \cite{agostinelli2003geant4}; adoption of GEANT4 for radiotherapy treatment verification however, is increasing \cite{agostinelli2003geant4,caccia2010medlinac2, jan2011gate, spezi2008overview, grevillot2011simulation, rodrigues2004application}.
Flexible geometry definition and physics process customisation provides the user with a high level of control, and the opportunity to simulate a wide range of radiotherapy techniques including brachytherapy, hadrontherapy and intensity modulated radiotherapy \cite{allison2006geant4}.
Significant computational overhead prevents the routine use of these Monte Carlo techniques in the clinical setting, however the advent of cloud computing provides a low cost and easy to maintain alternative to the set-up of dedicated compute hardware \cite{keyes2010radiation}, something that may be of particular benefit to clinics in rural and regional areas and developing countries.
Indeed, several authors have explored the usefulness of the cloud for Monte Carlo simulation \cite{gruntorad2010international, farbin2009emerging, silverman2010chep}, the most notable of which uses Fluka for proton beam dose calculations on the Amazon Elastic Compute Cloud (Amazon Web Services LLC, USA)\cite{keyes2010radiation}.

Amazon Web Services (AWS) provide organisations and individuals with the opportunity to leverage unused or under utilised Amazon network capacity for the purposes of scalable service provision such as high demand web-hosting with volatile loading conditions and scientific computation problems requiring significant compute or memory resources \cite{amazon2011ec2types}.
Under the AWS umbrella there are are number of specific services providing distinct capability, the most relevant of which for this study are discussed.
Amazon Elastic Compute Cloud (EC2) provides scalable compute through a number of predefined instances, where an instance is a virtual hardware device (or physical hardware device for select cases) with a predefined compute capability; the full gamut of instance types is outlined in table \ref{tab:ec2}.
Compute capability of a particular instance type is described using the EC2 compute unit, where one compute unit is the equivalent CPU capacity of a 1.0-1.2 GHz 2007 Opteron or 2007 Xeon processor \cite{amazon2011ec2types}.
At creation, any EC2 instance may have custom user data parsed to it, the user data itself may take on any form whether it be binary, ASCII or otherwise - subsequently this user data may be used to uniquely configure running tasks on the instances, or indeed the instance itself.
\begin{table*}
\begin{tabular}{llcccc}
Type	        & API Name                 & Compute Units & Processors         & RAM (GB) & Storage (GB)\\
\hline\hline
Standard Small     & \texttt{m1.small}     & 1             & 1 virtual          & 1.7      & 160\\
Standard Large     & \texttt{m1.large}     & 4             & 4 virtual          & 7.5      & 850\\
Standard 1X Large  & \texttt{m1.xlarge}    & 8             & 8 virtual          & 15       & 1690\\
Micro              & \texttt{t1.micro}     & 2 (burst)     & 2 virtual          & 0.613    & EBS\\
High Mem. 1X Large & \texttt{m2.xlarge}    & 6.5           & 2 virtual          & 17.1     & 420\\
High Mem. 2X Large & \texttt{m2.2xlarge}   & 13            & 4 virtual          & 34.2     & 850\\
High Mem. 4X Large & \texttt{m2.4xlarge}   & 26            & 8 virtual          & 68.4     & 1690\\
High CPU Medium    & \texttt{c1.medium}    & 5             & 2 virtual          & 1.7      & 350\\
High CPU 1X Large  & \texttt{c1.xlarge}    & 20            & 8 virtual          & 7        & 1690\\
Cluster 4X Large   & \texttt{cc1.4xlarge}  & 33.5          & 2 Intel Xeon X5570 & 23       & 1690\\
\hline
\end{tabular}
\caption{Types of preconfigured instance types available to the user on EC2 \cite{amazon2011ec2types}. The compute capability of the Micro type is burst only; the maximum compute cannot be sustained for lengthy periods. CPU bases instances are shown only; the Cluster GPU Quad Extra Large instance type is also available based on the Cluster 4X Large type with the addition of 2 NVIDIA Tesla Fermi M2050 GPU's \cite{amazon2011ec2types}.}
\label{tab:ec2} 
\end{table*}

In addition to EC2, AWS provides a redundant storage solution for the persistence of data.
Amazon Simple Storage Solution (S3) enables bulk upload and download of data associated with compute tasks, as well as provision for persisting the shutdown state of an instance - this is accomplished via the elastic block storage (EBS) virtual device which is backed by S3 \cite{amazon2011ebs}.
Access to the resources provided by EC2 and S3 can be performed programatically using the \texttt{boto} Python module \cite{garnaat2010boto} or directly via the AWS dashboard using a web browser.

Access to most services associated with AWS attract a usage fee \cite{amazon2011ec2pricing}.
Charges associated with S3 are at fixed rates where storage volume and events such as disk input/output are charged separately.
Three fee regimes are available for the user to select from when using the EC2 service.
On demand usage attracts a flat hourly rate dependant to instance type, and a secondary fee structure provides the opportunity for substantially reduced hourly usage rates with the payment of a yearly subscription in order to reserve a dedicated instance.
Dedicated instance reservation becomes increasingly economical as instance uptime and usage approaches 100\%  \cite{amazon2011ec2pricing}.
The third fee structure is delivered via a spot market where the user may enter the maximum bid price one is willing to pay for a given instance; price fluctuations of the spot market are governed by supply and demand on the market at the time.
If the spot price exceeds the maximum bid price for a running instance, the instance is automatically terminated.
Further, hourly rates are not prorated for partial instance hour usage; the hourly runtime of each instance is rounded up to the nearest hour.
Twenty instances running for half an hour each (10 hours of use) would be billed as 20 instance hours, whereas one instance running for 10 hours would be billed as only 10 instance hours for example.

Here within we describe in the process of executing a pre-existing GEANT4 simulation of a clinical linear accelerator \cite{cornelius2011commissioning} on the Amazon EC2 computing resource.
With a Python (Python Software Foundation, USA)\cite{rossum2011python} interface to the simulation, the \texttt{boto} Python module for AWS is used to distribute jobs in the cloud environment from the local user machine.

\section{Methods}
    \subsection{Clinical Linear Accelerator Simulation}
    A Varian Clinac was commissioned and calibrated for absolute dose calculation as described elsewhere\cite{cornelius2011commissioning}.
    A multi-step commissioning approach was used to tune the simulation so as to match depth dose and beam profile measurements in a water tank; the commissioning was carried out for a range of jaw defined field sizes using local compute resources.
    Further, the widely used intensity modulated radiation therapy (IMRT) verification test known as the chair test was simulated locally and compared to measurement \cite{cornelius2011commissioning}.
    All simulation calculations were verified with measurement using gamma evaluation and a pass/fail distance to agreement criterion of $3\%/3\ mm$ \cite{cornelius2011commissioning}.
    
    Using the \texttt{boost::python} C++ libraries\cite{abrahams2004boost} and the \texttt{g4py}\cite {murakami2006geant4} Python bindings already present in the GEANT4 toolkit, an interface to the linear accelerator simulation was created; Python interface examples distributed with the GEANT4 toolkit served as a template.
    Instantiating the Python class \texttt{Lianc} provided a basic linac set-up with default values for parameters such as MLC and jaw positions and gantry rotation.
    Property constructs with both get and set methods such as \texttt{Linac.energy} allowed for direct access to all parameters that define linac operation and simulation configuration. 
    Phantom geometry definition was also possible through the Python interface with \texttt{g4py} using standard techniques.
    
%
%
%
  
    \subsection{AWS Instance Set-up}
    A single instance of type \texttt{t1.micro} was launched using the pre-built and official Ubuntu 10.04 LTS 64 bit Amazon Machine Image (AMI) with identifier \texttt{ami-3202f25b}, booting from EBS.
    Elastic Block Storage was selected over the standard instance storage as EBS enables faster boot and persistence of data saved to disk after instance shutdown; however it should be noted that data saved to the instance disk would be lost on termination - distinct from shutdown as termination effectively destroys the instance \cite{amazon2011ebs}.
    The boot process itself was similar to the normal boot process for a default install of any recent version of the Ubuntu server distribution\cite{ubuntu2011ec2}.
    Unlike a conventional local install however, the \texttt{libcloud}\cite{libcloud2011online} package was installed by default on the AMI enabling access to instance user data parsed to the instance at the time of creation.
    Using a public/private key-pair generated using the AWS dashboard, remote access and administration of the instance was established using a secure shell (SSH); the fully qualified Dynamic Name Server (DNS) address of the instance was made available to the user through the AWS dashboard (right click on instance $\rightarrow$ \textit{Connect} menu item).
    GEANT4 version 9.3 and its dependencies were compiled and installed on the instance as well as other packages including \texttt{boost::python} and the \texttt{numpy} Numerical Python module \cite{ascher2010numericalv1.5}.
    Where available, pre-built binaries in the Ubuntu software repositories were favoured over compiling software from source.
    Once configured, the instance was saved as a custom and private AMI using the menu options available in the AWS dashboard (right click on instance $\rightarrow$ \textit{Save instance as AMI} menu item)- this custom AMI was then available to boot up to 20 instances with the default AWS account set-up.
    In the case of booting 20 High CPU Extra Large EC2 instances, 160 CPU cores were made available to the user with a total compute capability of 400 EC2 units.

    \subsection{Distributing Jobs in the Cloud}
\begin{figure*}
        \centering
        \includegraphics[]{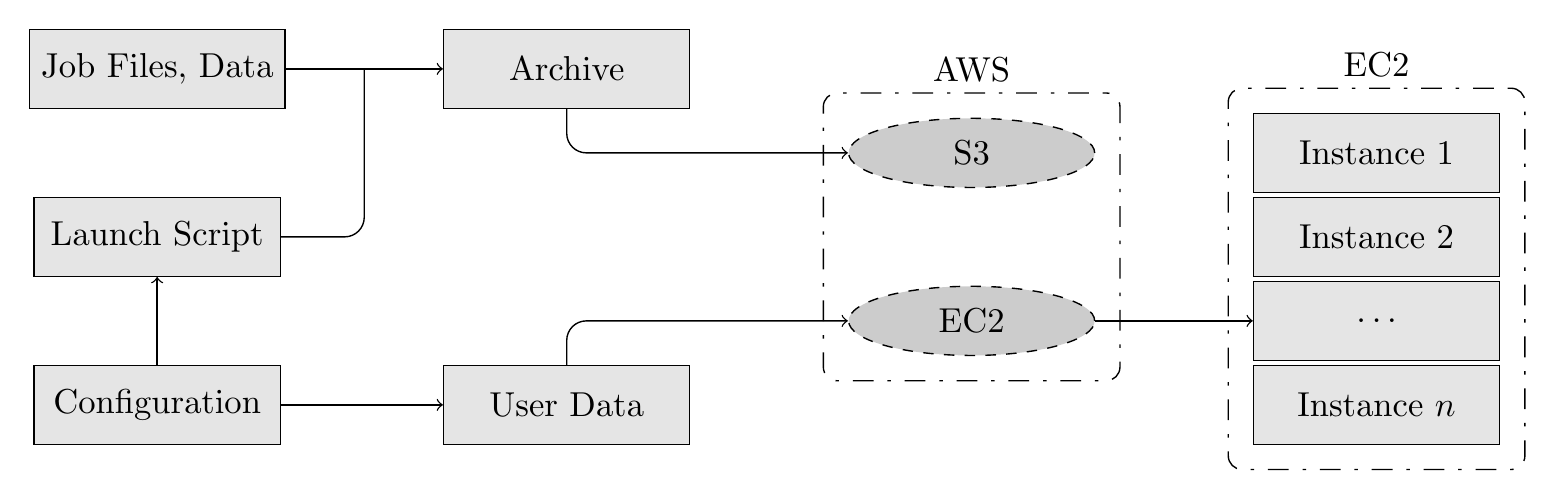}
        \caption{Launching EC2 instances from the local user machine.}
        \label{fig:flow_local}
    \end{figure*}

    Using \texttt{boto}, the Python API for AWS including EC2 and S3, a job launcher was created that managed the packing of a job description and data into a compressed archive and the launching of a group instances, see figure \ref{fig:flow_local}.
    For a given job, the simulation configuration included a manifest of all files and folders to be included as job data.
    Using the \texttt{tarfile} Python module, part of the Python standard library \cite{rossum2011python}, each file or folder in the manifest was added to an archive, followed by compression and writing to disk.
    From the local user machine, the compressed job archive was uploaded to S3 one time per unique simulation using \texttt{boto}.
    An EC2 reservation was requested which launched the prescribed number of instances for the job; a process fully managed by the \texttt{boto} Python module and EC2.
    Each instance had user data containing the simulation configuration including the location of the job archive on S3 transmitted to it automatically.
    
\begin{figure*}
        \centering
        \includegraphics[]{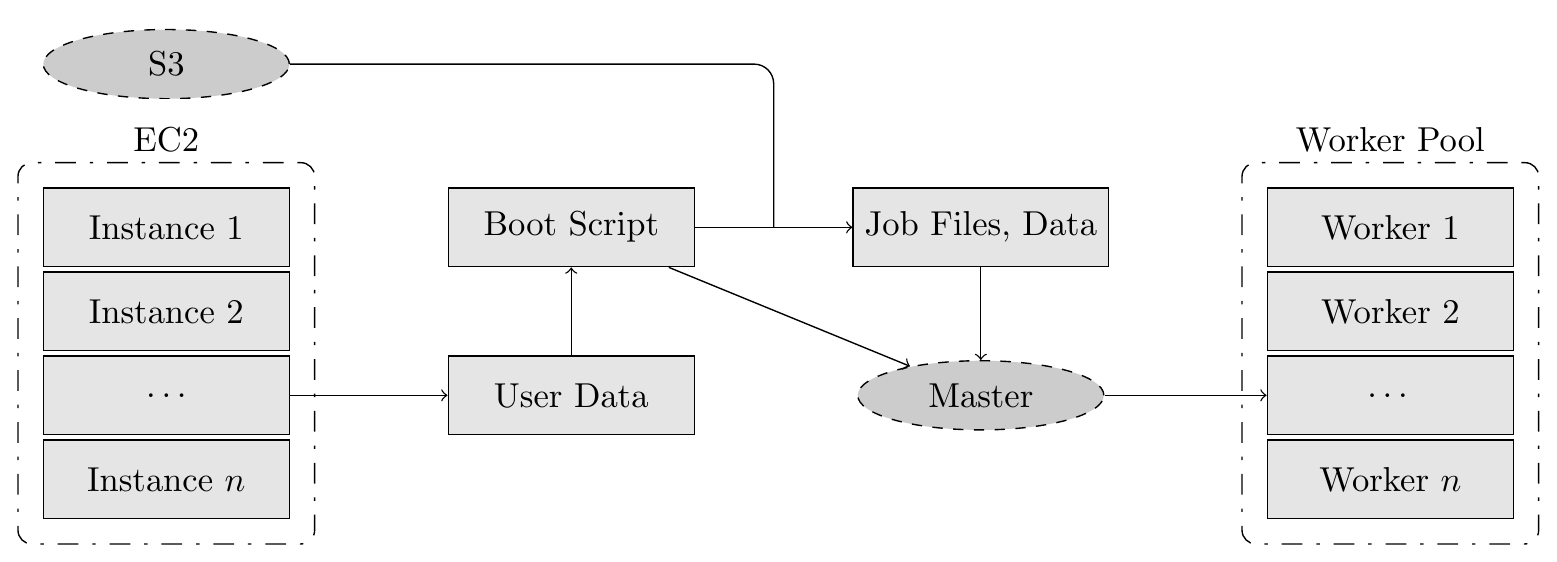}
        \caption{Simulation configuration and worker pool creation on each EC2 instance.}
        \label{fig:flow_instance}
    \end{figure*}

    At instance boot time, a Python script was automatically executed, recovering the simulation configuration from the pre-transmitted user data and launching a pool of worker processes with a pool size equal to the number of processor cores available on the instance, see figure \ref{fig:flow_instance}.
    The worker pool was created using the \texttt{multiprocessing} Python module \cite{rossum2011python}, again part of the Python standard library enabling a simulation described in Python function to be executed multiple times and concurrently across a number of processes equal to the pool size.
    On each instance, the master process managing the pool of worker processes waited for all workers to finish execution, subsequently combining and compressing the results returned by each worker process.
    
    Finally, the compressed result was uploaded to S3 to a location specified in the simulation configuration and the instance was terminated as soon as possible, thus minimising the potential of cost escalation.
    Retrieving results from S3 could be performed using the AWS dashboard and a web-browser.
    For execution of instances on the spot market, a maximum bid price could be specified at the time of reservation and configured as a parameter along with all other simulation parameters.
    From the user perspective, there was no difference between a instance acquired on-demand or bid for on the spot market.
    
    \subsection{Benchmarking Performance and Cost}
    \label{sub:benchmark}
    High CPU Extra Large EC2 instances were chosen for all jobs executed in the cloud as they provided the highest on-demand compute density per dollar, see section \ref{subsec:results_costs}.
    A series of test simulations were performed so as to examine simulation performance as a function of EC2 instance count.
    Using the GEANT4 geometry primitive \texttt{G4Box}, a $40\ cm$ cubic water phantom was defined and positioned with its center at the iso-center of the linear accelerator; $100\ cm$ source to axis distance (SAD) or $80\ cm$ source to surface distance (SSD).
    Irradiated with a jaw defined $5\times5\ cm$ field with gantry and primary collimator angles set to zero, $2.5\times10^{6}$ electrons incident on the copper target in the linear accelerator treatment head were simulated.
    The simulation was repeated for a range of EC2 instance counts ($1 \leq n \leq 20$) on the spot market (max price $= 0.30\ USD$)  with simulation completion time (the time elapsed from starting a job to uploading a result to S3), instance uptime, total simulation time (the total real CPU time used) and total simulation cost recorded.
    On-demand instance cost was calculated from the billed instance hours multiplied by the on-demand rate for the High CPU Extra Large instance type and compared to the actual cost incurred as a result of simulating the above using instances bid for on the spot market.
    Finally, historical data from January $1^{st}$ to April $18^{th}$ 2011 was acquired for each instance type using functionality provided by \texttt{boto} allowing for spot price history to be downloaded, and basic descriptive statistics were calculated.

\section{Results}
    \subsection{Simulation Output}
\begin{figure*}
      \subfloat[]{%
	    \begin{minipage}[c][1\width]{%
	       0.5\textwidth}
	       \centering%
	       \label{fig:depth_dose}\includegraphics[width=250px]{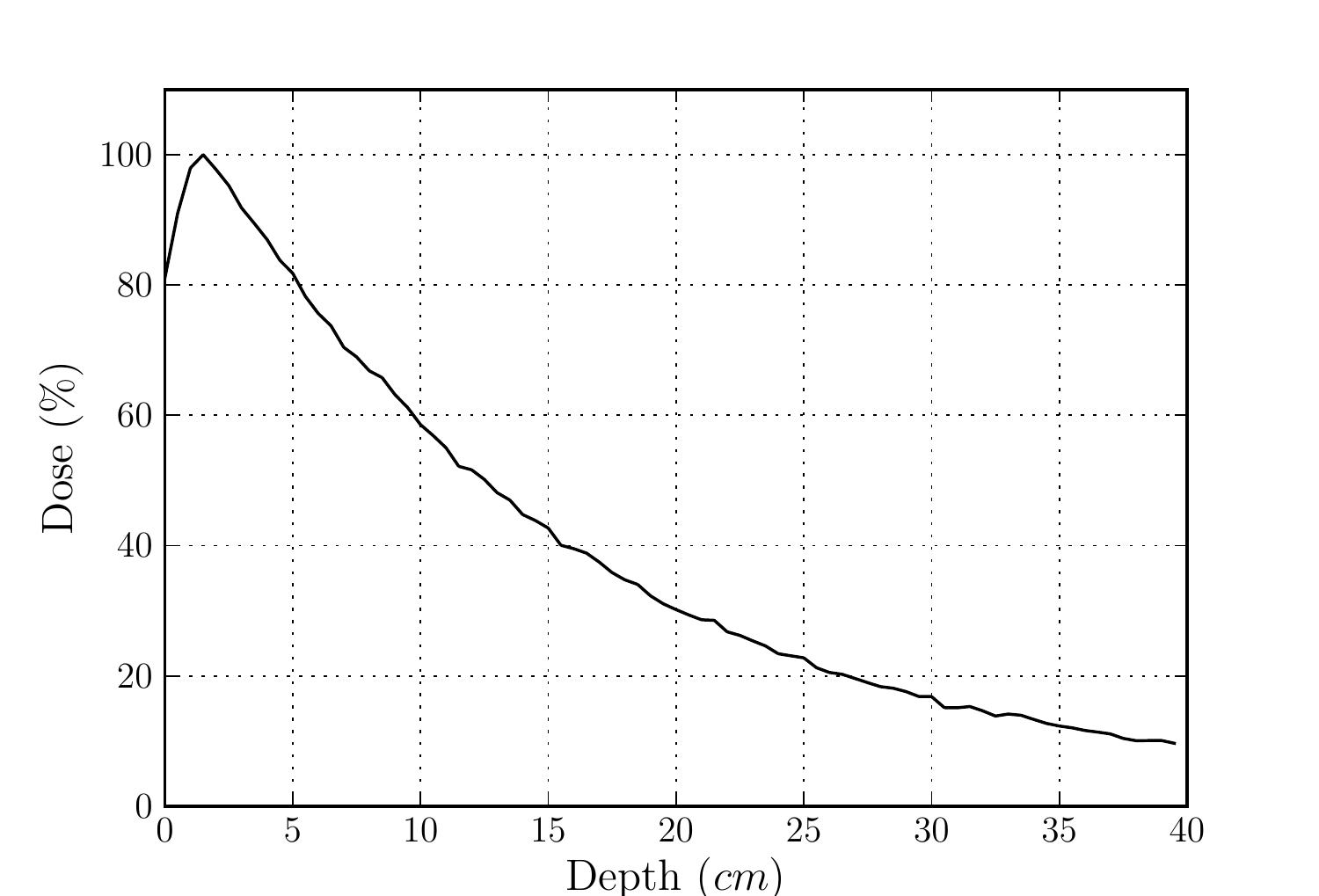}
	    \end{minipage}}
      \subfloat[]{%
	    \begin{minipage}[c][1\width]{%
	       0.5\textwidth}
	       \centering%
	       \label{fig:dose_dist}\includegraphics[width=250px]{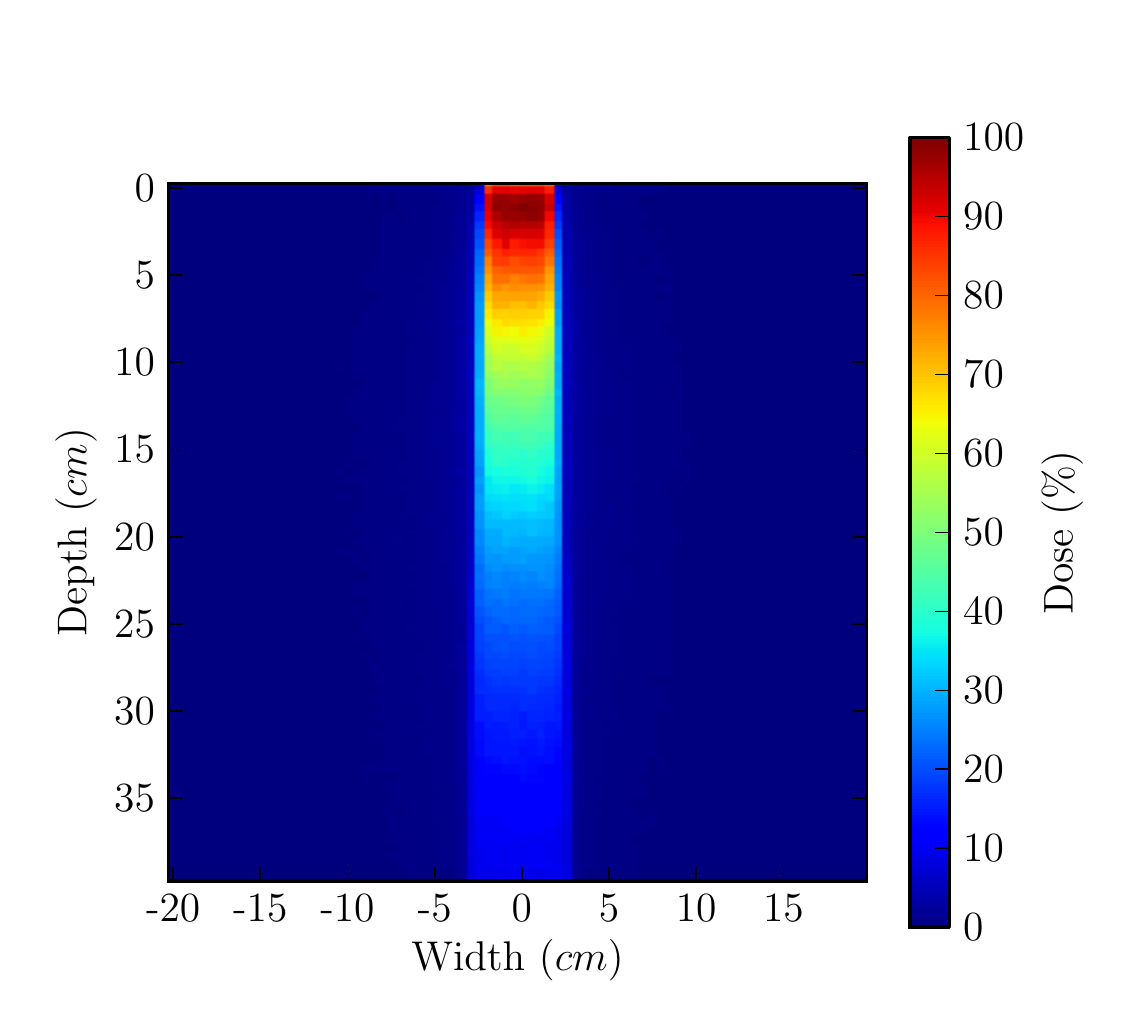}
	    \end{minipage}}
    \caption{Simulation output; \subref{fig:depth_dose} shows the central axis depth dose and \subref{fig:dose_dist} shows the dose distribution of the central slice in the water phantom. Note that the iso-center of the simulated linear accelerator was positioned at $(0,\ 20)$ in \subref{fig:dose_dist}.}
    \label{fig:dose}
    \end{figure*}

    Figure \ref{fig:dose} shows typical output for the simulation described in section \ref{sub:benchmark} using a $2\ mm$ scoring dose grid.
    All dose values are shown normalised to the maximum central axis dose.
    The size in memory for the entire dose grid  with $128\times128\times128$ voxels using single precision floating point values was $8\ MB$ per worker process for a total of $64\ MB$ per instance. 
    
    \subsection{Compute Performance}
\begin{figure*}
      \centering
      \subfloat[][]{\label{fig:compute}\includegraphics[width=250px]{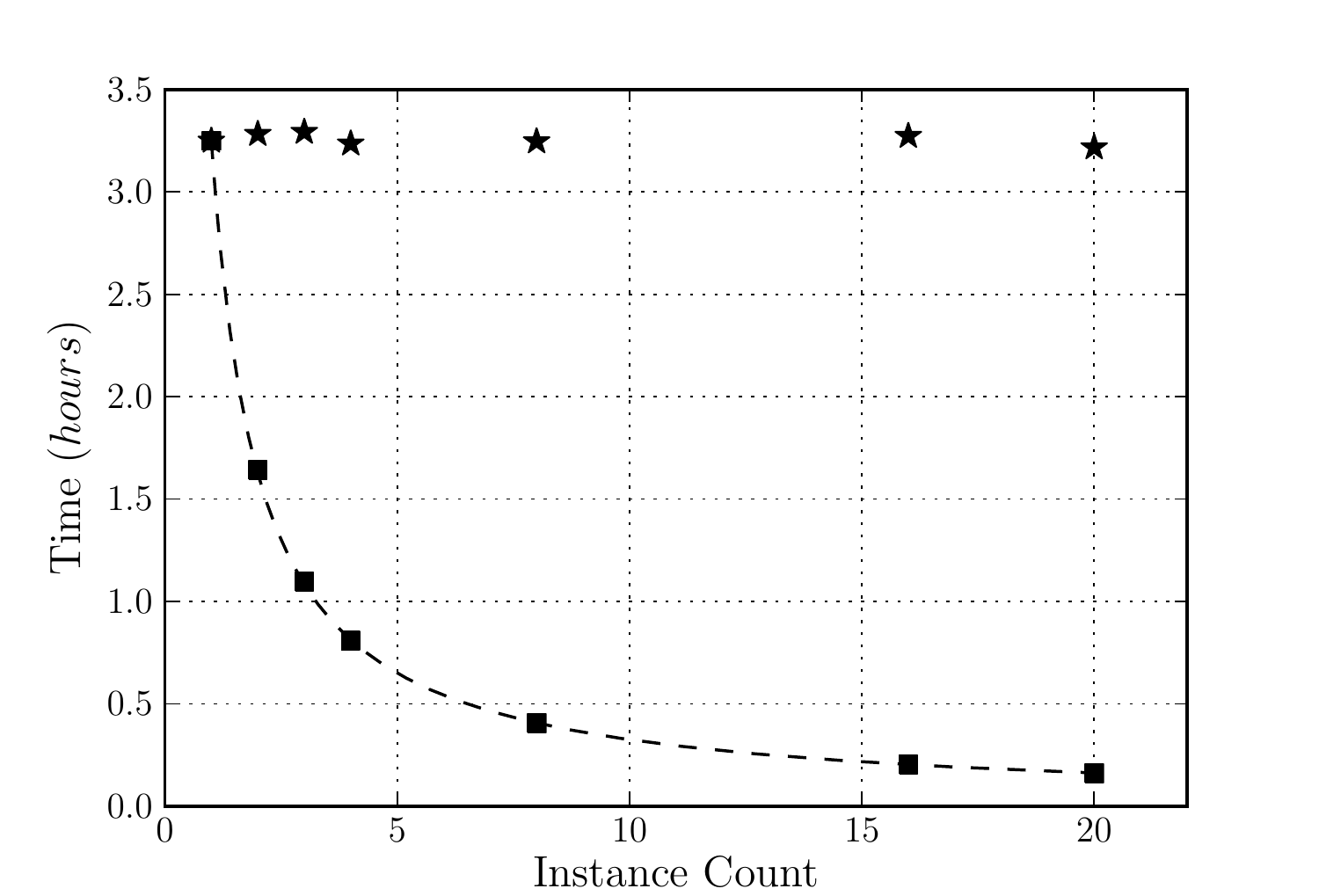}}                
      \subfloat[][]{\label{fig:cost}\includegraphics[width=250px]{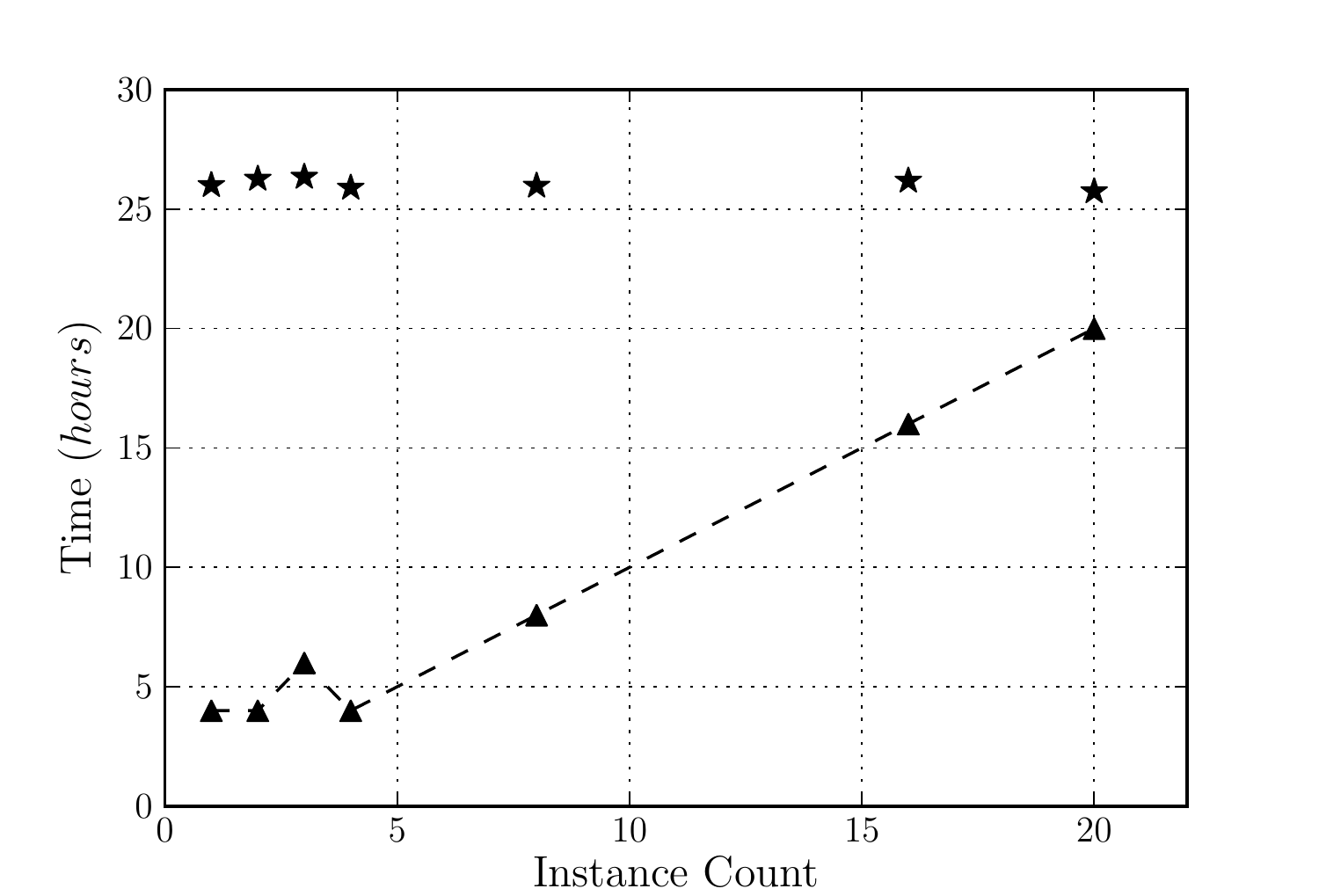}}

      \caption{Simulation time \subref{fig:compute} where $\bigstar$'s indicate the total instance up-time, $\blacksquare$'s indicate the time to simulation completion and the dashed line indicates the predicted simulation completion time (equation \ref{eq:time}). Billable instance time \subref{fig:cost} as a function of instance count where $\bigstar$'s indicate the total compute required, $ \blacktriangle$'s indicate the billable instance time, and the dashed line indicates the predicted billable instance time (equation \ref{eq:cost}).}
      \label{fig:graphs}
    \end{figure*}

    For the simulation described in section \ref{sub:benchmark} the average time from instance boot to the start of the simulation on the same node was $59\pm1 s$.
    Figure \ref{fig:graphs}\subref{fig:compute} shows the simulation completion time $t_{c}$ as a function of instance count; it was found to follow
    \begin{equation}\label{eq:time}
        t_{c} = \frac{t_{s}}{n_{i}n_{p}},
    \end{equation}
    where $t_{s}$ is the total simulation time required, $n_{i}\in\mathbb{N}^{\star} = \lbrace1,2,3,\ldots, 20\rbrace$ is the number of instances used per job and $n_{p}\in\mathbb{N}^{\star} = \lbrace1,2,3,\ldots, 8\rbrace$ is the number of processors available per instance.
    Noting that the default AWS accounts allowed for a maximum of $n_{i} = 20$ instances, and the maximum number of processors available per instances was $n_{p} = 8$ as of April 2011\cite{amazon2011ec2types}. Total simulation time or the total real CPU time consumed for the simulation as a function of instance count is shown in figure \ref{fig:graphs}\subref{fig:cost}.
    Mean total simulation time required for the simulation described in section \ref{sub:benchmark} was $t_{s}=26.1\pm0.2\ hours$ where the uncertainty represents one standard deviation about the mean.
    
    \subsection{Usage Costs}
    \label{subsec:results_costs}   
\begin{figure*} 
      \centering
        \includegraphics[width=250px]{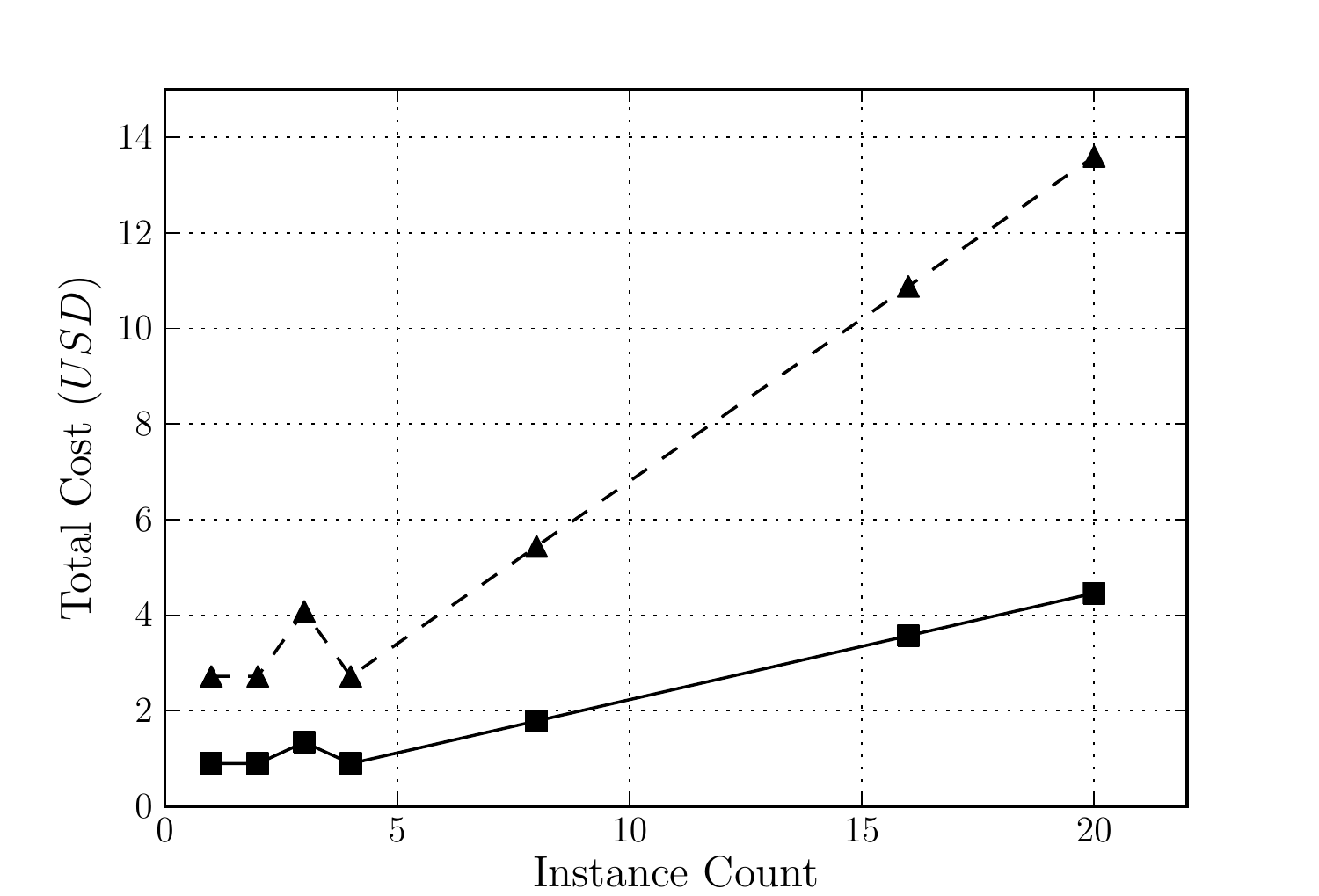}           
      \caption{Simulation cost as a function of instance count where $\blacksquare$'s indicate the incurred cost as a result of bidding for Amazon EC2 High CPU Extra Large instances on the spot market $(0.223\ USD/hour)$, $\blacktriangle$'s indicate the equivalent cost had the on-demand rate of $0.68\ USD/hour$ been charged, and the solid and dashed lines indicate the predicted instances hours (equation \ref{eq:cost}) multiplied by the hourly rate.}
      \label{fig:cost_compare}
    \end{figure*}

    Historical spot prices for the year 2011 to April $18^{th}$ for an Amazon EC2 High CPU Extra Large instance were acquired.
    A mean spot price of $0.34\pm0.13\ USD$ over this period was one half of the on-demand instance price ($0.68\ USD/hour$) - a general trend observed for most EC2 instance types, see table \ref{tab:ec2}.
    At the time of simulation, the quoted spot price for an Amazon EC2 High CPU Extra Large instance was $0.223\ USD/hour$, approximately one third of the on-demand instance.
     Where the instance count was greater than the simulation completion time in hours, cost escalation was linear with increasing instance count, see figure \ref{fig:cost_compare}.
    Billable instances hours required to complete a given job requiring $t_{s}$ total compute hours were found to follow
    \begin{equation}\label{eq:cost}
        t_{i} = n_{i}\left\lceil \frac{t_{s}}{n_{i}n_{p}} \right\rceil = n_{i}\left\lceil t_{c} \right\rceil,
    \end{equation}
     where $t_{i}\in\mathbb{N}^{\star} = \lbrace1,2,3,\ldots\rbrace$ is the total billable instance hours and $\lceil\ldots\rceil$ indicates the ceiling function, noting that the uptime of a given instance was rounded up to the nearest hour for the purposes of billing.
     Simulations running at least total cost were found where the simulation time in hours was wholly  divisible by the total number of instances running for that job, corresponding to the factors of $\lceil t_{s}/n_{p}\rceil\in\mathbb{N}^{\star} = \lbrace1,2,3,\ldots\rbrace$.

    \begin{table*}
        \begin{tabular}{lccccccc}
        \multicolumn{2}{c}{} & \multicolumn{2}{c}{Cost $(USD/hr)$}&& \multicolumn{2}{c}{Cost $(USD/hr/compute\ unit)$} \\
        \cline{3-4}\cline{6-7}
        API Name              & Compute Units & On-demand & Spot Average        & diff \%   & On-demand & Spot Average \\
        \hline\hline
        \texttt{m1.small}     & 1             & 0.085         & $0.043\pm0.009$ & 51         & 0.085 & 0.043          \\
        \texttt{m1.large}     & 4             & 0.34          & $0.16\pm0.05$   & 47         & 0.085 & 0.04 \\
        \texttt{m1.xlarge}    & 8             & 0.68          & $0.30\pm0.09$   & 45         & 0.085 & 0.038\\
        \texttt{t1.micro}     & 2 (burst)     & 0.02          & $0.012\pm0.003$ & 60         & 0.01  & 0.006\\
        \texttt{m2.xlarge}    & 6.5           & 0.50          & $0.21\pm0.07$   & 41         & 0.077 & 0.032\\
        \texttt{m2.2xlarge}   & 13            & 1.00          & $0.44\pm0.09$   & 44         & 0.077 & 0.034\\
        \texttt{m2.4xlarge}   & 26            & 2.00          & $0.87\pm0.17$   & 44         & 0.077 & 0.034\\
        \texttt{c1.medium}    & 5             & 0.17          & $0.087\pm0.03$  & 51         & 0.034 & 0.017\\
        \texttt{c1.xlarge}    & 20            & 0.68          & $0.34\pm0.13$   & 50         & 0.034 & 0.017\\
        \texttt{cc1.4xlarge}  & 33.5          & 1.60          & $0.57\pm0.06$   & 35         & 0.048 & 0.017\\
        \hline
        \end{tabular}
        \caption{Long-term spot market instance prices for the year 2011 to April $18^{th}$ for a range of preconfigured EC2 instance types. Uncertainty in the average spot price is one standard deviation about the mean.}
        \label{tab:price} 
    \end{table*}
    
\section{Discussion \& Conclusion}
    Using a GEANT4 simulation of a clinical linear accelerator, executed on the Amazon Elastic Compute Cloud, we have demonstrated the potential usefulness of cloud computing for rapid radiotherapy dose calculation.
    Additionally, a simple formulation allowing for the optimal selection of instance count for least cost has been proposed, given some estimate of total simulation time required.
    Figure \ref{fig:graphs}\subref{fig:compute} shows simulation time decreasing as $1/n$ with increasing instance count as observed by others \cite{keyes2010radiation}, cost however increases linearly with increasing instance count when simulation time in hours is less than the instance count, as shown in figure \ref{fig:graphs}\subref{fig:cost}.
    For a given simulation, if time is not a critical factor, the number of instances used can be tuned for least cost by ensuring each instance is in use for whole hours, as Amazon EC2 instances charges are not prorated for partial instance hour usage.
    However, in an environment where time is critical, increasing instance count reduces simulation time with a linearly increasing cost penalty.

    Two fee structures were examined when considering EC2 usage; a direct comparison between the actual cost inured as a result of using instances bid for on the spot market and the projected cost of acquiring the same instances had on-demand rates been charged.
    At the time of simulation, the spot market price of a single instance of type High CPU Extra Large was $0.223\ USD/hour$ and approximately one third of the on-demand price for the same instance type.
    This was less than the long term average of $0.34\pm0.13\ USD/hour$ at one half of the on-demand price; a general trend observed for all instance types.
    Whist volatility in the instance market may result in somewhat unpredictable expenditure, generally it is at least 50\% cheaper to use the instance spot market to acquire EC2 instances for computation.
   
   Application of this technique enables a GEANT4 user to perform a simulation in a distributed compute environment, with a low entry cost and no express need for dedicated compute hardware.
   For clinics in developing countries for example, which may not have sufficient resources to provide adequate cancer care\cite{hanna2010cancer} much less manage dedicated compute hardware, this may be of particular benefit.
   Indeed, the shortfall in the quality of cancer care in developing countries has been identified by others\cite{shakespeare2006external, hanna2010cancer}, in particular the relationship between inadequate staff training and suboptimal treatment delivery\cite{shakespeare2006external}.
   Systems to remedy this have been proposed by others, and of particular note is the Hospital Platform for E-health (HOPE) \cite{diarena2008hope} enabling the remote verification of radiotherapy treatment plans and other diagnostic and therapeutic tests.
   Adoption of initiatives such as HOPE, coupled with the computational resources provided by the cloud and the simulation techniques described here within may offer significant scientific and social benefit.
   
   Further work will explore any potential differences in dose calculations performed using local computing resources and resources that are provided by the cloud.
   Presently this work is part of a software toolkit using GEANT4 for the simulation of clinical linear accelerators\cite{cornelius2011commissioning}.
   Source code for running GEANT4 simulations on EC2 as described here within is freely available and may be obtained from: \texttt{http://code.google.com/p/manysim/}

\begin{acknowledgements}
This work is funded by the Queensland Cancer Physics Collaborative, and Cancer Australia (Department of Health and Ageing) Research Grant 614217.
\end{acknowledgements}

\nocite{*}


\begin{thebibliography}{10}%
\makeatletter
\providecommand \@ifxundefined [1]{%
 \ifx #1\undefined \expandafter \@firstoftwo
 \else \expandafter \@secondoftwo
\fi
}%
\providecommand \@ifnum [1]{%
 \ifnum #1\expandafter \@firstoftwo
 \else \expandafter \@secondoftwo
\fi
}%
\providecommand \enquote [1]{``#1''}%
\providecommand \bibnamefont  [1]{#1}%
\providecommand \bibfnamefont [1]{#1}%
\providecommand \citenamefont [1]{#1}%
\providecommand\href[0]{\@sanitize\@href}%
\providecommand\@href[1]{\endgroup\@@startlink{#1}\endgroup\@@href}%
\providecommand\@@href[1]{#1\@@endlink}%
\providecommand \@sanitize [0]{\begingroup\catcode`\&12\catcode`\#12\relax}%
\@ifxundefined \pdfoutput {\@firstoftwo}{%
 \@ifnum{\z@=\pdfoutput}{\@firstoftwo}{\@secondoftwo}%
}{%
 \providecommand\@@startlink[1]{\leavevmode}%
 \providecommand\@@endlink[0]{}%
}{%
 \providecommand\@@startlink[1]{%
  \leavevmode
  \pdfstartlink
   attr{/Border[0 0 1 ]/H/I/C[0 1 1]}%
   user{/Subtype/Link/A<</Type/Action/S/URI/URI(#1)>>}%
  \relax
 }%
 \providecommand\@@endlink[0]{\pdfendlink}%
}%
\providecommand \url  [0]{\begingroup\@sanitize \@url }%
\providecommand \@url [1]{\endgroup\@href {#1}{\urlprefix}}%
\providecommand \urlprefix [0]{URL }%
\providecommand \Eprint[0]{\href }%
\@ifxundefined \urlstyle {%
  \providecommand \doi [1]{doi:\discretionary{}{}{}#1}%
}{%
  \providecommand \doi [0]{doi:\discretionary{}{}{}\begingroup
  \urlstyle{rm}\Url }%
}%
\providecommand \doibase [0]{http://dx.doi.org/}%
\providecommand \Doi[1]{\href{\doibase#1}}%
\providecommand \selectlanguage [0]{\@gobble}%
\providecommand \bibinfo [0]{\@secondoftwo}%
\providecommand \bibfield [0]{\@secondoftwo}%
\providecommand \translation [1]{[#1]}%
\providecommand \BibitemOpen[0]{}%
\providecommand \bibitemStop [0]{}%
\providecommand \bibitemNoStop [0]{.\EOS\space}%
\providecommand \EOS [0]{\spacefactor3000\relax}%
\providecommand \BibitemShut [1]{\csname bibitem#1\endcsname}%
\bibitem{agostinelli2003geant4}%
  \BibitemOpen
  \bibfield{author}{%
  \bibinfo {author} {\bibfnamefont{S.}~\bibnamefont{Agostinelli}}, \bibinfo
  {author} {\bibfnamefont{J.}~\bibnamefont{Allison}}, \bibinfo {author}
  {\bibfnamefont{K.}~\bibnamefont{Amako}}, \bibinfo {author}
  {\bibfnamefont{J.}~\bibnamefont{Apostolakis}}, \bibinfo {author}
  {\bibfnamefont{H.}~\bibnamefont{Araujo}}, \bibinfo {author}
  {\bibfnamefont{P.}~\bibnamefont{Arce}}, \bibinfo {author}
  {\bibfnamefont{M.}~\bibnamefont{Asai}}, \bibinfo {author}
  {\bibfnamefont{D.}~\bibnamefont{Axen}}, \bibinfo {author}
  {\bibfnamefont{S.}~\bibnamefont{Banerjee}}, \bibinfo {author}
  {\bibfnamefont{G.}~\bibnamefont{Barrand}}, \emph{et~al.},\ }%
  \bibfield{title}{%
  \enquote{\bibinfo {title} {{Geant4 - a simulation toolkit}},}\ }%
  \bibfield{journal}{%
  \bibinfo {journal} {Nuclear Instruments and Methods in Physics
  Research-Section A Only}\ }%
  \textbf{\bibinfo {volume} {506}},\ \bibinfo {pages} {250--303} (\bibinfo
  {year} {2003})\BibitemShut{NoStop}%
\bibitem{caccia2010medlinac2}%
  \BibitemOpen
  \bibfield{author}{%
  \bibinfo {author} {\bibfnamefont{B.}~\bibnamefont{Caccia}}, \bibinfo {author}
  {\bibfnamefont{C.}~\bibnamefont{Andenna}},\ and\ \bibinfo {author}
  {\bibfnamefont{G.~A.~P.}\ \bibnamefont{Cirrone}},\ }%
  \bibfield{title}{%
  \enquote{\bibinfo {title} {{MedLinac2: a GEANT4 based software package for
  radiotherapy}},}\ }%
  \bibfield{journal}{%
  \bibinfo {journal} {Annali dell{'}Istituto superiore di sanit{\`{a}}}\ }%
  \textbf{\bibinfo {volume} {46}},\ \bibinfo {pages} {173--177} (\bibinfo
  {year} {2010})\BibitemShut{NoStop}%
\bibitem{jan2011gate}%
  \BibitemOpen
  \bibfield{author}{%
  \bibinfo {author} {\bibfnamefont{S.}~\bibnamefont{Jan}}, \bibinfo {author}
  {\bibfnamefont{D.}~\bibnamefont{Benoit}}, \bibinfo {author}
  {\bibfnamefont{E.}~\bibnamefont{Becheva}}, \bibinfo {author}
  {\bibfnamefont{T.}~\bibnamefont{Carlier}}, \bibinfo {author}
  {\bibfnamefont{F.}~\bibnamefont{Cassol}}, \bibinfo {author}
  {\bibfnamefont{P.}~\bibnamefont{Descourt}}, \bibinfo {author}
  {\bibfnamefont{T.}~\bibnamefont{Frisson}}, \bibinfo {author}
  {\bibfnamefont{L.}~\bibnamefont{Grevillot}}, \bibinfo {author}
  {\bibfnamefont{L.}~\bibnamefont{Guigues}}, \bibinfo {author}
  {\bibfnamefont{L.}~\bibnamefont{Maigne}}, \emph{et~al.},\ }%
  \bibfield{title}{%
  \enquote{\bibinfo {title} {{GATE V6: a major enhancement of the GATE
  simulation platform enabling modelling of CT and radiotherapy}},}\ }%
  \bibfield{journal}{%
  \bibinfo {journal} {Physics in Medicine and Biology}\ }%
  \textbf{\bibinfo {volume} {56}},\ \bibinfo {pages} {881} (\bibinfo {year}
  {2011})\BibitemShut{NoStop}%
\bibitem{spezi2008overview}%
  \BibitemOpen
  \bibfield{author}{%
  \bibinfo {author} {\bibfnamefont{E.}~\bibnamefont{Spezi}}\ and\ \bibinfo
  {author} {\bibfnamefont{G.}~\bibnamefont{Lewis}},\ }%
  \bibfield{title}{%
  \enquote{\bibinfo {title} {{An overview of Monte Carlo treatment planning for
  radiotherapy}},}\ }%
  \bibfield{journal}{%
  \bibinfo {journal} {Radiation protection dosimetry}}%
   (\bibinfo {year} {2008})\BibitemShut{NoStop}%
\bibitem{grevillot2011simulation}%
  \BibitemOpen
  \bibfield{author}{%
  \bibinfo {author} {\bibfnamefont{L.}~\bibnamefont{Grevillot}}, \bibinfo
  {author} {\bibfnamefont{T.}~\bibnamefont{Frisson}}, \bibinfo {author}
  {\bibfnamefont{D.}~\bibnamefont{Maneval}}, \bibinfo {author}
  {\bibfnamefont{N.}~\bibnamefont{Zahra}}, \bibinfo {author}
  {\bibfnamefont{J.~N.}\ \bibnamefont{Badel}},\ and\ \bibinfo {author}
  {\bibfnamefont{D.}~\bibnamefont{Sarrut}},\ }%
  \bibfield{title}{%
  \enquote{\bibinfo {title} {{Simulation of a 6 MV Elekta Precise Linac photon
  beam using GATE/GEANT4}},}\ }%
  \bibfield{journal}{%
  \bibinfo {journal} {Physics in Medicine and Biology}\ }%
  \textbf{\bibinfo {volume} {56}},\ \bibinfo {pages} {903} (\bibinfo {year}
  {2011})\BibitemShut{NoStop}%
\bibitem{rodrigues2004application}%
  \BibitemOpen
  \bibfield{author}{%
  \bibinfo {author} {\bibfnamefont{P.}~\bibnamefont{Rodrigues}}, \bibinfo
  {author} {\bibfnamefont{A.}~\bibnamefont{Trindade}}, \bibinfo {author}
  {\bibfnamefont{L.}~\bibnamefont{Peralta}}, \bibinfo {author}
  {\bibfnamefont{C.}~\bibnamefont{Alves}}, \bibinfo {author}
  {\bibfnamefont{A.}~\bibnamefont{Chaves}},\ and\ \bibinfo {author}
  {\bibfnamefont{M.~C.}\ \bibnamefont{Lopes}},\ }%
  \bibfield{title}{%
  \enquote{\bibinfo {title} {{Application of GEANT4 radiation transport toolkit
  to dose calculations in anthropomorphic phantoms}},}\ }%
  \bibfield{journal}{%
  \bibinfo {journal} {Applied Radiation and Isotopes}\ }%
  \textbf{\bibinfo {volume} {61}},\ \bibinfo {pages} {1451--1461} (\bibinfo
  {year} {2004})\BibitemShut{NoStop}%
\bibitem{allison2006geant4}%
  \BibitemOpen
  \bibfield{author}{%
  \bibinfo {author} {\bibfnamefont{J.}~\bibnamefont{Allison}}, \bibinfo
  {author} {\bibfnamefont{K.}~\bibnamefont{Amako}}, \bibinfo {author}
  {\bibfnamefont{J.}~\bibnamefont{Apostolakis}}, \bibinfo {author}
  {\bibfnamefont{H.}~\bibnamefont{Araujo}}, \bibinfo {author}
  {\bibfnamefont{P.A.}\ \bibnamefont{Dubois}}, \bibinfo {author}
  {\bibfnamefont{M.}~\bibnamefont{Asai}}, \bibinfo {author}
  {\bibfnamefont{G.}~\bibnamefont{Barrand}}, \bibinfo {author}
  {\bibfnamefont{R.}~\bibnamefont{Capra}}, \bibinfo {author}
  {\bibfnamefont{S.}~\bibnamefont{Chauvie}}, \bibinfo {author}
  {\bibfnamefont{R.}~\bibnamefont{Chytracek}}, \emph{et~al.},\ }%
  \bibfield{title}{%
  \enquote{\bibinfo {title} {{Geant4 developments and applications}},}\ }%
  \bibfield{journal}{%
  \bibinfo {journal} {Nuclear Science, IEEE Transactions on}\ }%
  \textbf{\bibinfo {volume} {53}},\ \bibinfo {pages} {270--278} (\bibinfo
  {year} {2006})\BibitemShut{NoStop}%
\bibitem{keyes2010radiation}%
  \BibitemOpen
  \bibfield{author}{%
  \bibinfo {author} {\bibfnamefont{R.W.}\ \bibnamefont{Keyes}}, \bibinfo
  {author} {\bibfnamefont{C.}~\bibnamefont{Romano}}, \bibinfo {author}
  {\bibfnamefont{D.}~\bibnamefont{Arnold}},\ and\ \bibinfo {author}
  {\bibfnamefont{S.}~\bibnamefont{Luan}},\ }%
  \bibfield{title}{%
  \enquote{\bibinfo {title} {{Radiation therapy calculations using an on-demand
  virtual cluster via cloud computing}},}\ }%
  \bibfield{journal}{%
  \bibinfo {journal} {Arxiv preprint arXiv:1009.5282}}%
   (\bibinfo {year} {2010})\BibitemShut{NoStop}%
\bibitem{gruntorad2010international}%
  \BibitemOpen
  \bibfield{author}{%
  \bibinfo {author} {\bibfnamefont{J.}~\bibnamefont{Gruntorad}}\ and\ \bibinfo
  {author} {\bibfnamefont{M.}~\bibnamefont{Lokajicek}},\ }%
  \enquote{\bibinfo {title} {{International Conference on Computing in High
  Energy and Nuclear Physics (CHEP'09)}},}\ in\ \emph{\bibinfo {booktitle}
  {Journal of Physics: Conference Series}},\ Vol.\ \bibinfo {volume} {219}\
  (\bibinfo {year} {2010})\ p.\ \bibinfo {pages} {001001}\BibitemShut{NoStop}%
\bibitem{farbin2009emerging}%
  \BibitemOpen
  \bibfield{author}{%
  \bibinfo {author} {\bibfnamefont{A.}~\bibnamefont{Farbin}},\ }%
  \bibfield{title}{%
  \enquote{\bibinfo {title} {{Emerging Computing Technologies in High Energy
  Physics}},}\ }%
  \bibfield{journal}{%
  \bibinfo {journal} {Arxiv preprint arXiv:0910.3440}}%
   (\bibinfo {year} {2009})\BibitemShut{NoStop}%
\bibitem{silverman2010chep}%
  \BibitemOpen
  \bibfield{author}{%
  \bibinfo {author} {\bibfnamefont{A}~\bibnamefont{Silverman}}, \bibinfo
  {author} {\bibfnamefont{I}~\bibnamefont{Fedorko}}, \bibinfo {author}
  {\bibfnamefont{W}~\bibnamefont{Lapka}},\ and\ \bibinfo {author}
  {\bibfnamefont{G}~\bibnamefont{Lo~Presti}},\ }%
  \enquote{\bibinfo {title} {{CHEP 2010 Report. CHEP - Computing in High Energy
  and nuclear Physics}},}\ \bibinfo {type} {Tech. Rep.}\ \bibinfo {number}
  {CERN-IT-Note-2010-007}\ (\bibinfo {institution} {CERN},\ \bibinfo {address}
  {Geneva},\ \bibinfo {year} {2010})\BibitemShut{NoStop}%
\bibitem{amazon2011ec2types}%
  \BibitemOpen
  \enquote{\bibinfo {title} {{Amazon EC2 Instance Types}},}\ \bibinfo
  {howpublished} {Amazon Web Services LLC.
  http://aws.amazon.com/ec2/instance-types/} (\bibinfo {month} {April}\
  \bibinfo {year} {2011})\BibitemShut{NoStop}%
\bibitem{amazon2011ebs}%
  \BibitemOpen
  \enquote{\bibinfo {title} {{Amazon EBS}},}\ \bibinfo {howpublished} {Amazon
  Web Services LLC. http://aws.amazon.com/ebs/} (\bibinfo {month} {April}\
  \bibinfo {year} {2011})\BibitemShut{NoStop}%
\bibitem{garnaat2010boto}%
  \BibitemOpen
  \bibfield{author}{%
  \bibinfo {author} {\bibfnamefont{M.}~\bibnamefont{Garnaat}} \emph{et~al.},\
  }%
  \emph{\bibinfo {title} {{Boto Python interface to Amazon Web Services
  Documentation}}},\ \bibinfo {organization} {Computer software.
  http://code.google.com/p/boto/},\ \bibinfo {edition} {v2.0}\ ed. (\bibinfo
  {year} {2010})\BibitemShut{NoStop}%
\bibitem{amazon2011ec2pricing}%
  \BibitemOpen
  \enquote{\bibinfo {title} {{Amazon EC2 Pricing}},}\ \bibinfo {howpublished}
  {Amazon Web Services LLC. http://aws.amazon.com/ec2/pricing/} (\bibinfo
  {month} {April}\ \bibinfo {year} {2011})\BibitemShut{NoStop}%
\bibitem{cornelius2011commissioning}%
  \BibitemOpen
  \bibfield{author}{%
  \bibinfo {author} {\bibfnamefont{I.}~\bibnamefont{{Cornelius}}}, \bibinfo
  {author} {\bibfnamefont{B.}~\bibnamefont{{Hill}}}, \bibinfo {author}
  {\bibfnamefont{N.}~\bibnamefont{{Middlebrook}}}, \bibinfo {author}
  {\bibfnamefont{C.}~\bibnamefont{{Poole}}}, \bibinfo {author}
  {\bibfnamefont{B.}~\bibnamefont{{Oborn}}},\ and\ \bibinfo {author}
  {\bibfnamefont{C.}~\bibnamefont{{Langton}}},\ }%
  \bibfield{title}{%
  \enquote{\bibinfo {title} {{Commissioning of a Geant4 based treatment plan
  simulation tool: linac model and DICOM-RT interface}},}\ }%
  \bibfield{journal}{%
  \bibinfo {journal} {Arxiv preprint arXiv:1104.5082}}%
   (\bibinfo {month} {Apr.}\ \bibinfo {year} {2011})\BibitemShut{NoStop}%
\bibitem{rossum2011python}%
  \BibitemOpen
  \bibfield{author}{%
  \bibinfo {author} {\bibfnamefont{G.}~\bibnamefont{van Rossum}}\ and\ \bibinfo
  {author} {\bibfnamefont{F.~L.}\ \bibnamefont{Drake}},\ }%
  \emph{\bibinfo {title} {Python Reference Manual}},\ \bibinfo {organization}
  {Python Software Foundation. http://python.org/},\ \bibinfo {edition}
  {v2.7.1}\ ed. (\bibinfo {month} {April}\ \bibinfo {year}
  {2011})\BibitemShut{NoStop}%
\bibitem{abrahams2004boost}%
  \BibitemOpen
  \bibfield{author}{%
  \bibinfo {author} {\bibfnamefont{D.}~\bibnamefont{Abrahams}}, \bibinfo
  {author} {\bibfnamefont{U.}~\bibnamefont{Koethe}}, \bibinfo {author}
  {\bibfnamefont{RW}~\bibnamefont{Grosse-Kunstleve}}, \emph{et~al.},\ }%
  \emph{\bibinfo {title} {{The Boost Python Library Documentation}}},\ \bibinfo
  {organization} {Computer software . http://boost.org/libs/python/},\ \bibinfo
  {edition} {v1.41}\ ed. (\bibinfo {month} {October}\ \bibinfo {year}
  {2004})\BibitemShut{NoStop}%
\bibitem{murakami2006geant4}%
  \BibitemOpen
  \bibfield{author}{%
  \bibinfo {author} {\bibfnamefont{K.}~\bibnamefont{Murakami}}\ and\ \bibinfo
  {author} {\bibfnamefont{H.}~\bibnamefont{Yoshida}},\ }%
  \enquote{\bibinfo {title} {{A Geant4-Python Interface: Development and Its
  Applications}},}\ in\ \emph{\bibinfo {booktitle} {Nuclear Science Symposium
  Conference Record, 2006. IEEE}},\ Vol.~\bibinfo {volume} {1}\ (\bibinfo
  {organization} {IEEE},\ \bibinfo {year} {2006})\ pp.\ \bibinfo {pages}
  {98--100}\BibitemShut{NoStop}%
\bibitem{ubuntu2011ec2}%
  \BibitemOpen
  \bibinfo {organization} {Ubuntu Community Documentation.
  https://help.ubuntu.com/community/EC2StartersGuide},\ \emph{\bibinfo {title}
  {{Ubuntu EC2 Starters Guide}}} (\bibinfo {month} {March}\ \bibinfo {year}
  {2011})\BibitemShut{NoStop}%
\bibitem{libcloud2011online}%
  \BibitemOpen
  \bibinfo {organization} {Computer software.
  http://ci.apache.org/projects/libcloud/apidocs/},\ \emph{\bibinfo {title}
  {{Libcloud: a unified interface to the cloud}}},\ \bibinfo {edition}
  {v0.4.2}\ ed. (\bibinfo {year} {2011})\BibitemShut{NoStop}%
\bibitem{ascher2010numericalv1.5}%
  \BibitemOpen
  \bibfield{author}{%
  \bibinfo {author} {\bibfnamefont{D.}~\bibnamefont{Ascher}}, \bibinfo {author}
  {\bibfnamefont{P.F.}\ \bibnamefont{Dubois}}, \bibinfo {author}
  {\bibfnamefont{K.}~\bibnamefont{Hinsen}}, \bibinfo {author}
  {\bibfnamefont{J.}~\bibnamefont{Hugunin}}, \bibinfo {author}
  {\bibfnamefont{T.}~\bibnamefont{Oliphant}}, \emph{et~al.},\ }%
  \emph{\bibinfo {title} {{Numerical Python Documentation}}},\ \bibinfo
  {organization} {Computer software. http://numpy.scipy.org/},\ \bibinfo
  {edition} {v1.5}\ ed.\BibitemShut{Stop}%
\bibitem{hanna2010cancer}%
  \BibitemOpen
  \bibfield{author}{%
  \bibinfo {author} {\bibfnamefont{T.P.}\ \bibnamefont{Hanna}}\ and\ \bibinfo
  {author} {\bibfnamefont{A.C.T.}\ \bibnamefont{Kangolle}},\ }%
  \bibfield{title}{%
  \enquote{\bibinfo {title} {{Cancer control in developing countries: using
  health data and health services research to measure and improve access,
  quality and efficiency}},}\ }%
  \bibfield{journal}{%
  \bibinfo {journal} {BMC International Health and Human Rights}\ }%
  \textbf{\bibinfo {volume} {10}},\ \bibinfo {pages} {24} (\bibinfo {year}
  {2010}),\ ISSN \bibinfo {issn} {1472-698X}\BibitemShut{NoStop}%
\bibitem{shakespeare2006external}%
  \BibitemOpen
  \bibfield{author}{%
  \bibinfo {author} {\bibfnamefont{T.P.}\ \bibnamefont{Shakespeare}}, \bibinfo
  {author} {\bibfnamefont{M.F.}\ \bibnamefont{Back}}, \bibinfo {author}
  {\bibfnamefont{J.J.}\ \bibnamefont{Lu}}, \bibinfo {author}
  {\bibfnamefont{K.M.}\ \bibnamefont{Lee}},\ and\ \bibinfo {author}
  {\bibfnamefont{R.K.}\ \bibnamefont{Mukherjee}},\ }%
  \bibfield{title}{%
  \enquote{\bibinfo {title} {{External audit of clinical practice and medical
  decision making in a new Asian oncology center: results and implications for
  both developing and developed nations}},}\ }%
  \bibfield{journal}{%
  \bibinfo {journal} {International Journal of Radiation Oncology* Biology*
  Physics}\ }%
  \textbf{\bibinfo {volume} {64}},\ \bibinfo {pages} {941--947} (\bibinfo
  {year} {2006})\BibitemShut{NoStop}%
\bibitem{diarena2008hope}%
  \BibitemOpen
  \bibfield{author}{%
  \bibinfo {author} {\bibfnamefont{M.}~\bibnamefont{Diarena}}, \bibinfo
  {author} {\bibfnamefont{S.}~\bibnamefont{Nowak}}, \bibinfo {author}
  {\bibfnamefont{JY}~\bibnamefont{Boire}}, \bibinfo {author}
  {\bibfnamefont{V.}~\bibnamefont{Bloch}}, \bibinfo {author}
  {\bibfnamefont{D.}~\bibnamefont{Donnarieix}}, \bibinfo {author}
  {\bibfnamefont{A.}~\bibnamefont{Fessy}}, \bibinfo {author}
  {\bibfnamefont{B.}~\bibnamefont{Grenier}}, \bibinfo {author}
  {\bibfnamefont{B.}~\bibnamefont{Irrthum}}, \bibinfo {author}
  {\bibfnamefont{Y.}~\bibnamefont{Legr{\'e}}}, \bibinfo {author}
  {\bibfnamefont{L.}~\bibnamefont{Maigne}}, \emph{et~al.},\ }%
  \bibfield{title}{%
  \enquote{\bibinfo {title} {{HOPE, an open platform for medical data
  management on the grid.}}.}\ }%
  \bibfield{journal}{%
  \bibinfo {journal} {Studies in health technology and informatics}\ }%
  \textbf{\bibinfo {volume} {138}},\ \bibinfo {pages} {34} (\bibinfo {year}
  {2008})\BibitemShut{NoStop}%
\end{thebibliography}
\end{document}